\documentclass[aps, prd, floatfix, nofootinbib, showpacs, twocolumn, superscriptaddress]{revtex4-1}

\usepackage{amsmath,amsfonts,amssymb,bm}
\usepackage{graphicx}
\usepackage{color}
\usepackage{subfigure}
\usepackage{multirow}
\usepackage{textcomp}
\usepackage{slashed}

\usepackage{enumitem}

\usepackage[mathscr,scaled=1.15]{urwchancal}
\DeclareFontFamily{OT1}{pzc}{}
\DeclareFontShape{OT1}{pzc}{m}{it}%
{<-> s * [1.15] pzcmi7t}{}
\DeclareMathAlphabet{\mathpzc}{OT1}{pzc}{m}{it}

\definecolor{purple}{rgb}{0.5,0,0.5}
\definecolor{blue}{rgb}{0.0,0,0.9}
\definecolor{prdblue}{rgb}{0.133,0.118,0.498}
\usepackage[colorlinks=true, pdfstartview=FitV, linkcolor=prdblue, citecolor= prdblue, urlcolor=prdblue]{hyperref}

\begin{document}

\title{Mass-dependence of pseudoscalar meson elastic form factors}

\author{Muyang Chen}
\affiliation{School of Physics, Nankai University, Tianjin 300071, China}

\author{Minghui Ding}
\affiliation{School of Physics, Nankai University, Tianjin 300071, China}
\affiliation{Physics Division, Argonne National Laboratory, Argonne, Illinois
60439, USA}

\author{Lei Chang}
\email[]{leichang@nankai.edu.cn}
\affiliation{School of Physics, Nankai University, Tianjin 300071, China}

\author{Craig D.~Roberts}
\email[]{cdroberts@anl.gov}
\affiliation{Physics Division, Argonne National Laboratory, Argonne,
Illinois 60439, USA}

\date{27 August 2018}

\begin{abstract}
A continuum approach to quark-antiquark bound-states is used to determine the electromagnetic form factors of pion-like mesons with masses $m_{0^-}/{\rm GeV}=0.14, 0.47, 0.69, 0.83$ on a spacelike domain that extends to $Q^2 \lesssim 10\,$GeV$^2$.  The results enable direct comparisons with contemporary lattice-QCD calculations of heavy-pion form factors at large values of momentum transfer and aid in understanding them.  They also reveal, \emph{inter alia}, that the form factor of the physical pion provides the best opportunity for verification of the factorised hard-scattering formula relevant to this class of exclusive processes and that this capacity diminishes steadily as the meson mass increases.
\end{abstract}

\maketitle

\noindent\textbf{1.$\;$Introduction}.
Perturbation theory in quantum chromodynamics [QCD] is applicable to hard exclusive processes; and for almost forty years the leading-order factorised result for the electromagnetic form factor of a pseudoscalar meson has excited experimental and theoretical interest.  Namely \cite{Farrar:1979aw, Lepage:1979zb, Efremov:1979qk, Lepage:1980fj}, $\exists \, Q_0>\Lambda_{\rm QCD}$ such that
\begin{equation}
\label{EqHardScattering}
Q^2 F_{0^-}(Q^2) \stackrel{Q^2 > Q_0^2}{\approx} 16 \pi \alpha_s(Q^2)  f_{0^-}^2 \mathpzc{w}_{0^-}^2(Q^2),
\end{equation}
where: $f_{0^-}$ is the meson's leptonic decay constant; $\alpha_s(Q^2) $ is the leading-order strong running-coupling
\begin{equation}
\alpha_s(Q^2) = 4 \pi/[\beta_0\,\ln(Q^2/\Lambda^2_{\rm QCD})],
\label{alphaS}
\end{equation}
with $\beta_0 = 11 - (2/3) n_f$ [$n_f$ is the number of active quark flavours]; and
\begin{equation}
\label{wphi}
\mathpzc{w}_{0^-}(Q^2) = \frac{1}{3} \int_0^1 dx\, \frac{1}{x} \,\varphi_{0^-}(x;Q^2)\,,
\end{equation}
where $\varphi_{0^-}(x;Q^2)$ is the meson's dressed-valence-quark parton distribution amplitude [PDA].  This PDA is determined by the meson's light-front wave function and relates to the probability that, with constituents collinear up to the scale $\zeta=\surd Q^2$, a valence-quark within the meson carries light-front fraction $x$ of the bound-state's total momentum.  Here, $\Lambda_{\rm QCD} \sim 0.2\,$GeV is the empirical mass-scale of QCD.  

Crucially, the value of $Q_0$ is not predicted by perturbative QCD; but a hope that this scale might be as low as the proton mass, $m_p$, was influential in ensuring that the continuous electron beam accelerator facility [CEBAF] was planned with a peak beam energy of $4\,$GeV.

Before CEBAF began operations in 1994, the charged-pion elastic electromagnetic form factor, $F_\pi(Q^2)$, was only known on $Q^2\in[0,0.25]\,$GeV$^2$.  Measured by scattering high-energy pions from atomic electrons \cite{Dally:1981ur, Dally:1982zk, Amendolia:1984nz, Amendolia:1986wj}, this data yielded a sound measurement of the pion's charge radius.  
%
Owing to kinematic limitations on the energy of the pion beam and unfavorable momentum transfer, however, different experimental methods are required to reach higher $Q^2$.  Electroproduction of pions from the proton can serve this purpose; and in 1997 a long-planned CEBAF experiment collected data on $0.6 \leq Q^2/{\rm GeV}^2\leq 1.6$ \cite{Volmer:2000ek}.  Analyses of subsequent experiments, which reached $Q^2=2.45\,$GeV$^2$ by capitalising on higher beam energies available at a CEBAF exceeding original expectations, are described in Refs.\,\cite{Horn:2006tm, Horn:2007ug, Huber:2008id, Blok:2008jy}.  However, no signal for the behaviour in Eq.\,\eqref{EqHardScattering} has yet been claimed.  Consequently, experiments planned and approved at the upgraded Jefferson Lab [JLab\,12] aim for precision measurements of $F_\pi(Q^2)$ to $Q^2=6\,$GeV$^2$ and have the potential to reach $Q^2\approx 8.5\,$GeV$^2$ \cite{E12-06-101, E12-07-105, Horn:2017csb}.

Eq.\,\eqref{EqHardScattering} involves the meson's PDA, which is an essentially nonperturbative quantity.  Lacking reliable predictions for the pointwise form of $\varphi_\pi(x;Q^2)$ appropriate to existing experimental scales, original expectations for empirical values of $Q^2 F_\pi(Q^2)$ that would confirm Eq.\,\eqref{EqHardScattering} were based on the conformal limit result \cite{Lepage:1979zb, Efremov:1979qk, Lepage:1980fj}
\begin{equation}
\label{PDAcl}
\varphi_\pi(x;Q^2) \stackrel{\Lambda_{\rm QCD}^2/Q^2 \simeq 0}{\approx} \varphi^{\rm cl}(x) = 6 x (1-x)\,,
\end{equation}
in which case
\begin{equation}
\label{pionUV4}
Q^2 F_\pi(Q^2) \stackrel{Q^2=4\,{\rm GeV}^2}{\approx} 0.15\,.
\end{equation}
This prediction is a factor of $2.7$ smaller than the empirical value quoted at $Q^2 =2.45\,$GeV$^2$ \cite{Huber:2008id}: $0.41^{+0.04}_{-0.03}$.

Recently, however, continuum and lattice-QCD [lQCD] studies of the pseudoscalar meson bound-state problem have revealed that $\varphi_\pi(x;Q^2\sim (2m_p)^2)$ is a concave function, much broader than $\varphi^{\rm cl}(x)$ owing to emergent mass generation in the Standard Model \cite{Mikhailov:1986be, Petrov:1998kg, Brodsky:2006uqa, Arthur:2010xf, Chang:2013pq, Cloet:2013tta, Segovia:2013eca, Braun:2015axa, Horn:2016rip, Zhang:2017bzy}.  Using this information, a continuum calculation of $F_\pi(Q^2)$ on a large domain of spacelike momenta predicted \cite{Chang:2013nia} that the approved JLab\,12 experiments \cite{E12-06-101, E1207105} are capable of validating Eq.\,\eqref{EqHardScattering} because the estimate in Eq.\,\eqref{pionUV4} is too small by a factor of approximately two.

lQCD validation of this prediction would be welcome.  However, owing to competing demands [\emph{e.g}.\ large lattice volume to represent light pions, small lattice spacing to reach large $Q^2$, and high statistics to compensate for decaying signal-to-noise ratio as form factors drop rapidly with increasing $Q^2$], lQCD results with pion masses near the physical value, $m_\pi$, are currently restricted to small-$Q^2$: $0<Q^2\lesssim 0.25\,$GeV$^2$ \cite{Koponen:2015tkr, Alexandrou:2017blh}.  Such analyses provide information about the pion's charge radius, but do not address the questions of whether and at which scale Eq.\,\eqref{EqHardScattering} is empirically applicable.  No lQCD predictions at $m_\pi$ are available on the full domain accessible to JLab\,12, but new results exist on $Q^2 \lesssim 6\,$GeV$^2$ at bound-state mass-squared values $m_{0^-}^2 \approx 10m_\pi^2, 25 m_\pi^2$ \cite{Chambers:2017tuf, Koponen:2017fvm}.   Herein, employing the continuum approach to the QCD bound-state problem that was used \cite{Chang:2013nia} to calculate the pion form factor and reconcile its behaviour with Eq.\,\eqref{EqHardScattering}, we discuss how these modern lQCD results bear on validation of this hard scattering formula and related issues.

\smallskip

\noindent\textbf{2.$\;$Computing pseudoscalar meson form factors}.
At leading order in the systematic, symmetry-preserving Dyson-Schwinger equation [DSE] approximation scheme described in Refs.\,\cite{Munczek:1994zz, Bender:1996bb, Roberts:1996jxS}, \emph{viz}.\ rainbow-ladder [RL] truncation, the elastic form factor of a pion-like system constituted from degenerate current-quarks is given by \cite{Roberts:1994hh, Maris:1998hc, Maris:2000sk, Holl:2005vu, Bhagwat:2006pu}
\begin{eqnarray}
\nonumber
K_\mu F_{0^-}(Q^2) & = & N_c {\rm tr} 
\int\! \frac{d^4 k}{(2\pi)^4}\,
\chi_\mu(k+p_f,k+p_i) \\
&& \times \Gamma_{0^-}(k_i;p_i)\,S(k)\,\Gamma_{0^-}(k_f;-p_f)\,, \quad\label{RLFpi}
\end{eqnarray}
where $Q$ is the incoming photon momentum, $p_{f,i} = K\pm Q/2$, $k_{f,i}=k+p_{f,i}/2$, $p_{f,i}^2 = -m_{0^-}^2$, and the trace is over spinor indices.
The other elements in Eq.\,\eqref{RLFpi} are the dressed-quark propagator, $S(p)$,
which, consistent with Eq.\,\eqref{RLFpi}, is computed with the rainbow-truncation gap equation; and the $0^-$-meson Bethe-Salpeter amplitude $\Gamma_{0^-}(k;P)$
%
and unamputated dressed-quark-photon vertex, $\chi_\mu(k_f,k_i)$, both computed in RL truncation. 
[The impact of corrections to the RL computation is understood \cite{Raya:2015gva, Raya:2016yuj}.  The dominant effect is a modification of the power associated with the logarithmic running in Eq.\,\eqref{EqHardScattering}.  That running is slow and immaterial to the present discussion; but its effect can readily be incorporated when important.]

The leading-order DSE result for the pseudoscalar meson form factor is now determined once an interaction kernel is specified for the RL Bethe-Salpeter equation.  We use that explained in Ref.\,\cite{Qin:2011dd, Qin:2011xq}:
 \begin{subequations}
\label{KDinteraction}
\begin{align}
\mathscr{K}_{\alpha_1\alpha_1',\alpha_2\alpha_2'} & = {\mathpzc G}_{\mu\nu}(k) [i\gamma_\mu]_{\alpha_1\alpha_1'} [i\gamma_\nu]_{\alpha_2\alpha_2'}\,,\\
 {\mathpzc G}_{\mu\nu}(k) & = \tilde{\mathpzc G}(k^2) T_{\mu\nu}(k)\,,
\end{align}
\end{subequations}
with $k^2 T_{\mu\nu}(k) = k^2 \delta_{\mu\nu} - k_\mu k_\nu$ and ($s=k^2$)
\begin{align}
\label{defcalG}
 \tfrac{1}{Z_2^2}\tilde{\mathpzc G}(s) & =
 \frac{8\pi^2}{\omega^4} D e^{-s/\omega^2} + \frac{8\pi^2 \gamma_m \mathcal{F}(s)}{\ln\big[ \tau+(1+s/\Lambda_{\rm QCD}^2)^2 \big]}\,,
\end{align}
where $\gamma_m=4/\beta_0$, $\Lambda_{\rm QCD}=0.234\,$GeV, $\tau={\rm e}^2-1$, and ${\cal F}(s) = \{1 - \exp(-s/[4 m_t^2])\}/s$, $m_t=0.5\,$GeV.  $Z_2$ is the dressed-quark wave function renormalisation constant.
We employ a mass-independent momentum-subtraction renormalisation scheme for the gap and inhomogeneous vertex equations, implemented by making use of the scalar Ward-Green-Takahashi identity and fixing all renormalisation constants in the chiral limit \cite{Chang:2008ec}, with renormalisation scale $\zeta=2\,$GeV$=:\zeta_2$.  

The development of Eqs.\,\eqref{KDinteraction}, \eqref{defcalG} is summarised in Ref.\,\cite{Qin:2011dd} and their connection with QCD is described in Ref.\,\cite{Binosi:2014aea}; but it is worth reiterating some points.  For instance, the interaction is deliberately consistent with that determined in studies of QCD's gauge sector, which indicate that the gluon propagator is a bounded, regular function of spacelike momenta that achieves its maximum value on this domain at $s=0$ \cite{Bowman:2004jm, Boucaud:2011ug, Ayala:2012pb, Aguilar:2012rz, Binosi:2014aea, Binosi:2016xxu, Binosi:2016nme, Gao:2017uox, Rodriguez-Quintero:2018wma}, and the dressed-quark-gluon vertex does not possess any structure which can qualitatively alter these features \cite{Skullerud:2003qu, Bhagwat:2004kj, Aguilar:2014lha, Williams:2015cvx, Binosi:2016rxz, Binosi:2016wcx, Aguilar:2016lbe, Bermudez:2017bpx, Cyrol:2017ewj}.
It is specified in Landau gauge because, \emph{e.g}.\ this gauge is a fixed point of the renormalisation group and ensures that sensitivity to differences between \emph{Ans\"atze} for the gluon-quark vertex are least noticeable, thus providing the conditions for which rainbow-ladder truncation is most accurate.  
The interaction also preserves the one-loop renormalisation group behaviour of QCD so that, \emph{e.g}.\ the quark mass-functions produced are independent of the renormalisation point.
On the other hand, in the infrared, \emph{i.e}.\ $s \lesssim m_p^2$, Eq.\,\eqref{defcalG} defines a two-parameter model, the details of which determine whether confinement and/or dynamical chiral symmetry breaking [DCSB] are realised in solutions of the dressed-quark gap equations.

Computations \cite{Qin:2011dd, Qin:2011xq} reveal that many properties of light-quark ground-state vector- and isospin-nonzero pseudoscalar-mesons are practically insensitive to variations of $\omega \in [0.4,0.6]\,$GeV, so long as
\begin{equation}
 \varsigma^3 := D\omega = {\rm constant}.
\label{Dwconstant}
\end{equation}
This feature also extends to numerous characteristics of the nucleon and $\Delta$-baryon \cite{Eichmann:2008ef, Eichmann:2012zz}.  The value of $\varsigma$ is chosen to reproduce, as well as possible, the measured value of the pion's mass and leptonic decay constant; and in RL truncation this requires
\begin{equation}
\label{varsigmalight}
\varsigma  =0.82\,{\rm GeV}\,,
\end{equation}
with renormalisation-group-invariant current-quark mass
\begin{equation}
\hat m_u = \hat m_d = \hat m = 6.6\,{\rm MeV}\,,
\end{equation}
which corresponds to a one-loop evolved mass of $m^{\zeta_2} = 4.6\,$MeV.  We will subsequently employ $\omega=0.5\,$GeV, the midpoint of the insensitivity domain, and typically report the response of results to a 20\% variation in this value.

\begin{table*}[t!]
\caption{\label{TabResults}
Input current-quark masses [one-loop evolved from an associated value of $\hat m$] for four pion-like mesons and related results computed with $\omega=0.5\pm0.1\,$GeV in Eqs.\,\eqref{defcalG}-\eqref{varsigmalight}.  $\langle \xi^2\rangle$, $\alpha$ are defined in Eqs.\,\eqref{Eqxi2}, \eqref{Eqalpha}.
Empirically \cite{Olive:2016xmw}:
$f_\pi = 0.092\,$GeV,
$r_\pi = 0.672(8)\,$fm.
Regarding Row~2, the lQCD results at $m_{0^-}=0.47\,$GeV \cite{Chambers:2017tuf} are associated with $f_{0^-}=0.111(2)\,GeV$ \cite{Bornyakov:2016dzn}, $r_{0^-} = 0.56(1)\,$fm [our estimate, using monopole fit to lattice results]; and concerning Row~3, Ref.\,\cite{Koponen:2017fvm} reports $f_{0^-}=0.128\,$GeV, $r_{0^-} = 0.498(4)\,$fm for $m_{0^-}=0.69\,$GeV.
[In the table, all dimensioned quantities listed in GeV, except $r_{0^-}$, in fm.]}
\begin{tabular}{c|c|c|c|c|c|c|c|c|c|c|c|c|c}
\hline \hline
\multirow{2}{*}{$\;\;\;m^{\zeta_2}\;\;\;$}&
\multirow{2}{*}{\; $m_{0^-}\;$}	&
\multicolumn{4}{|c|}{$\omega=0.4$\;}&
\multicolumn{4}{|c|}{$\omega=0.5$\;}&
\multicolumn{4}{|c}{$\omega=0.6$\;} 	\\ [0.5mm]
\cline{3-14}
	&	& \; $f_{0^-}^{}$\; & $r_{0^-}$ &\;$\langle \xi^2 \rangle$\; &\; $\alpha$
&\;
$f_{0^-}^{}$\; & $r_{0^-}$ &\;$\langle \xi^2 \rangle$\; &\; $\alpha$
&\; $f_{0^-}^{}$\; & $r_{0^-}$ &\;$\langle \xi^2 \rangle$\; &\; $\alpha$\\
\hline
0.0046	& 0.14 & 0.092 &0.63 & 0.255 & 0.46 & 0.094 & 0.66 & 0.265 & 0.39 & 0.097 & 0.68 &0.273& 0.33\\
0.053\phantom{6}	& 0.47	& 0.115 & 0.53 &  0.217& 0.80 & 0.115 & 0.55 & 0.226 & 0.71 & 0.115 & 0.56 & 0.229	& 0.68\\
$0.107\phantom{6}$	& 0.69 & 0.135 & 0.47 &  0.196 & 1.05 & 0.133 & 0.49 & 0.207 & 0.92 & 0.133 & 0.49 & 0.211& 0.87 	\\
0.152\phantom{6}	& 0.83 & 0.147 & 0.43 &  0.180 & 1.28 & 0.145& 0.45 & 0.193 & 1.09 & 0.145 & 0.45 & 0.200& 1.00	\\
\hline \hline
\end{tabular}
\end{table*}

The RL approximation to the elastic electromagnetic form factor of a pion-like pseudoscalar meson with mass $m_{0^-}$ is now obtained as follows.
(i) Perform a coupled solution of the dressed-quark gap-  and meson Bethe-Salpeter-equations, defined via Eqs.\,\eqref{KDinteraction}, \eqref{defcalG}, varying the gap equation's current-quark mass until the Bethe-Salpeter equation has a solution at $P^2= - m_{0^-}^2$, following Ref.\,\cite{Maris:1997tm} and adapting the algorithm improvements from Ref.\,\cite{Krassnigg:2009gd} when necessary.
(ii) With the dressed-quark propagator obtained thereby and the same interaction, solve the inhomogeneous Bethe-Salpeter equation to obtain the unamputated dressed-quark-photon vertex, including its dependence on $Q^2$, as described, \emph{e.g}.\ in Ref.\,\cite{Maris:1999bh}.
(iii) Combine these elements to form the integrand in Eq.\,\eqref{RLFpi} and compute the integral as a function of $Q^2$ to extract the form factor, $F_{0^-}(Q^2)$; an exercise first completed in Ref.\,\cite{Maris:2000sk}.

To connect the results thus obtained and Eq.\,\eqref{EqHardScattering}, the associated meson PDA at the same renormalisation scale is needed.  It can be obtained from the meson's Poincar\'e-covariant Bethe-Salpeter amplitude following the methods described in Refs.\,\cite{Chang:2013pq, Cloet:2013tta, Segovia:2013eca, Li:2016dzv}.  Namely, one computes the leading non-trivial Mellin moment of the PDA via
\begin{align}
n\cdot P &f_{0^-}  \langle \xi^2 \rangle = 3 \, {\rm tr} Z_2\int \frac{d^4 k}{(2\pi)^4} \left[\frac{2 n\cdot k}{n\cdot P}\right]^2 \nonumber \\
& \times \gamma_{5} \gamma\cdot n\,S(k+P/2) \Gamma_{0^-}(k;P) S(k-P/2)\,, \label{Eqxi2}
\end{align}
with $\xi = (2 x -1)$, $P^2=-m_{0^-}^2$, $n^2=0$, $n\cdot P = -m_{0^-}$, using the same Poincar\'e-covariant regularisation of the integral as in the bound-state equations.  A convergence-factor $1/[1+k^2 r^2]$ is included in the integrand to stabilise the computation; the moment is computed as a function of $r^2$; and the final value is obtained by extrapolation to $r^2=0$.  This procedure is efficient and reliable \cite{Li:2016dzv}.
Using this moment,
which is zero when evaluated with $\varphi^{\rm cl}$, one can reconstruct a realistic approximation to the PDA by writing
\begin{equation}
\label{Eqalpha}
\varphi_{0^-}(x;\zeta_2) = x^\alpha (1-x)^\alpha \, \Gamma(2 [\alpha+1]) /\Gamma(\alpha+1)^2\,,
\end{equation}
with $\alpha$ chosen to reproduce the calculated value of $\xi^2$.  [The error in this procedure is negligible compared with that deriving from a 20\% variation of $\omega$ in Eq.\,\eqref{Dwconstant}.]

\smallskip

\begin{figure*}[!t]
\centering
\includegraphics[width=0.45\textwidth]{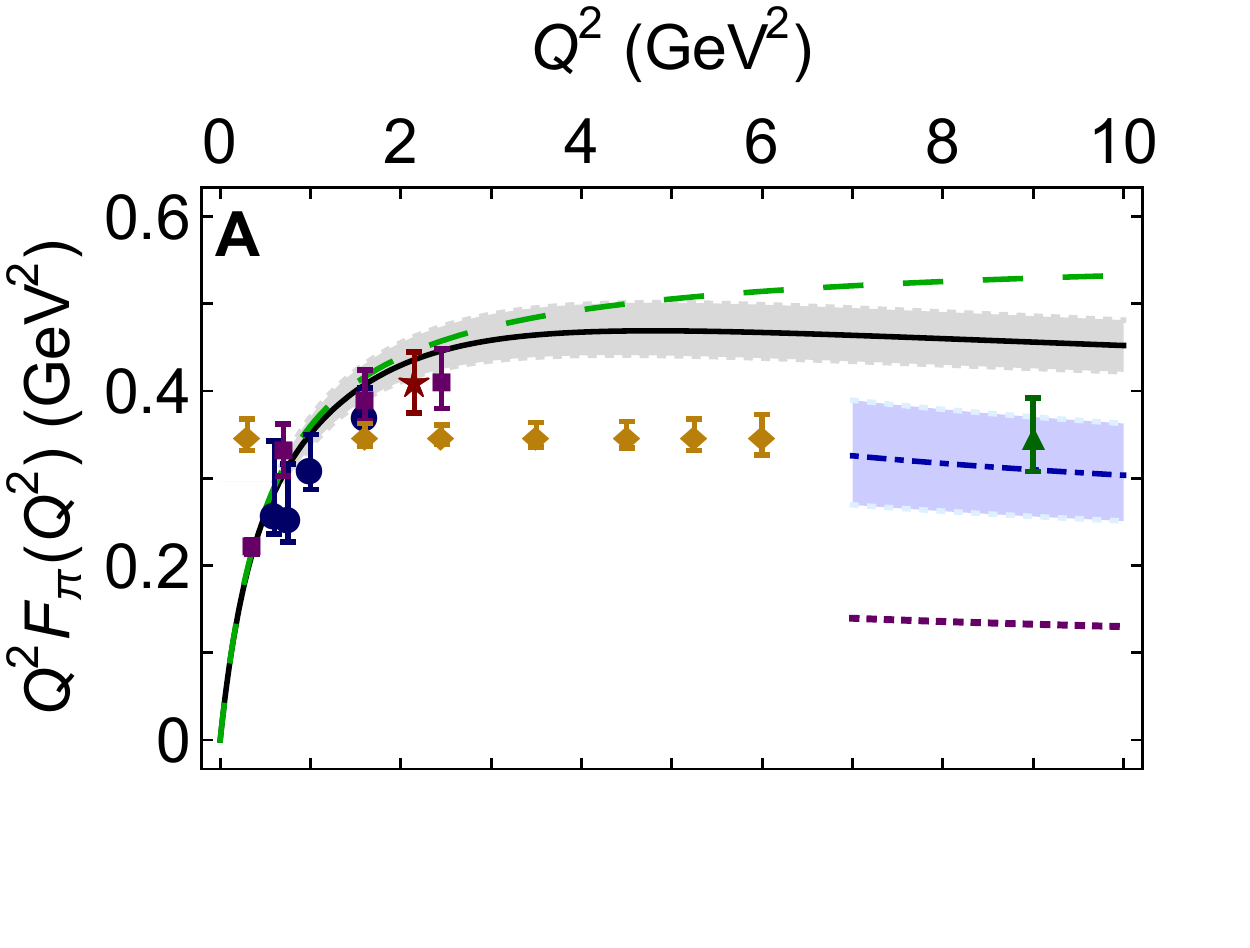}
\includegraphics[width=0.45\textwidth]{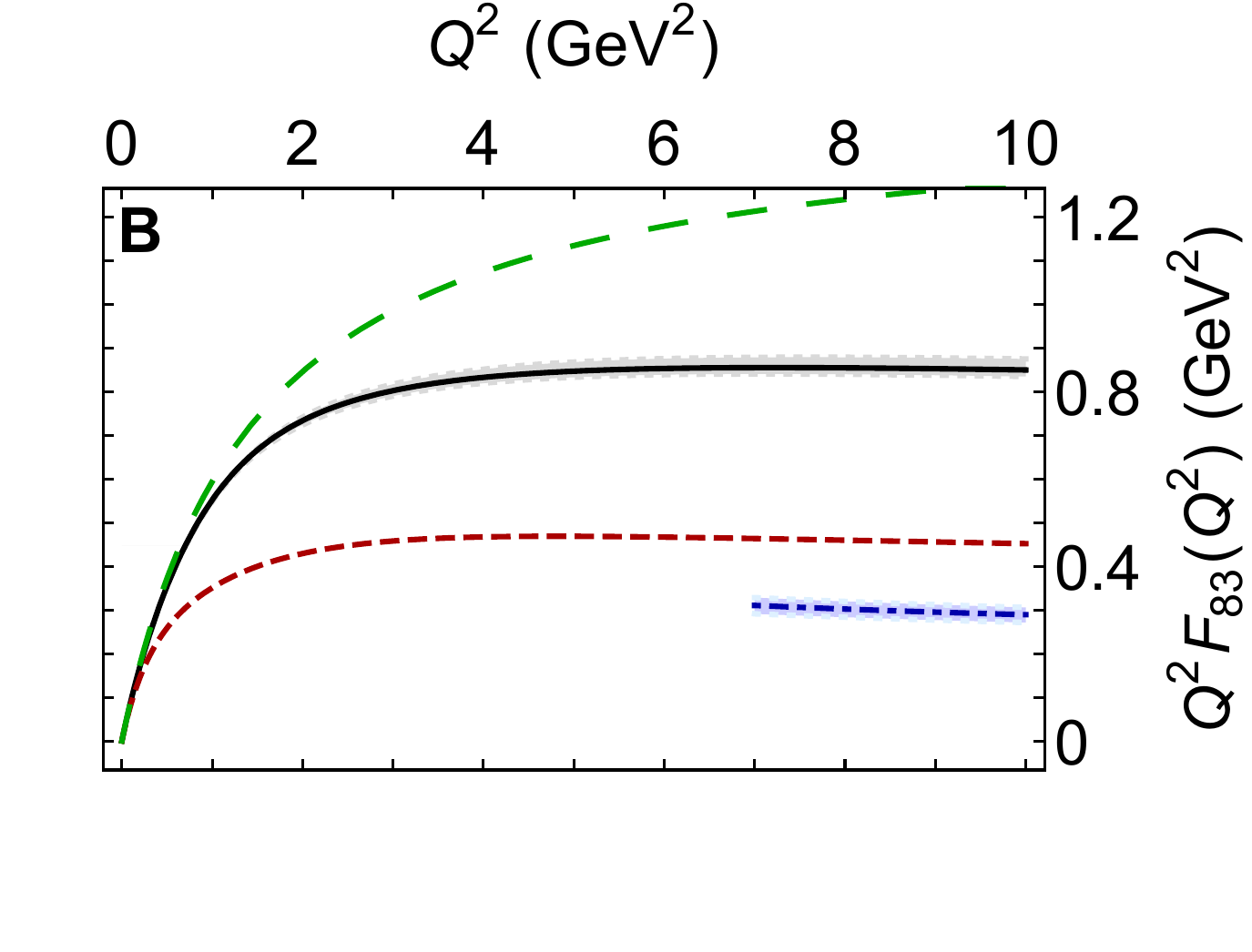}

\vspace*{-8ex}

\includegraphics[width=0.45\textwidth]{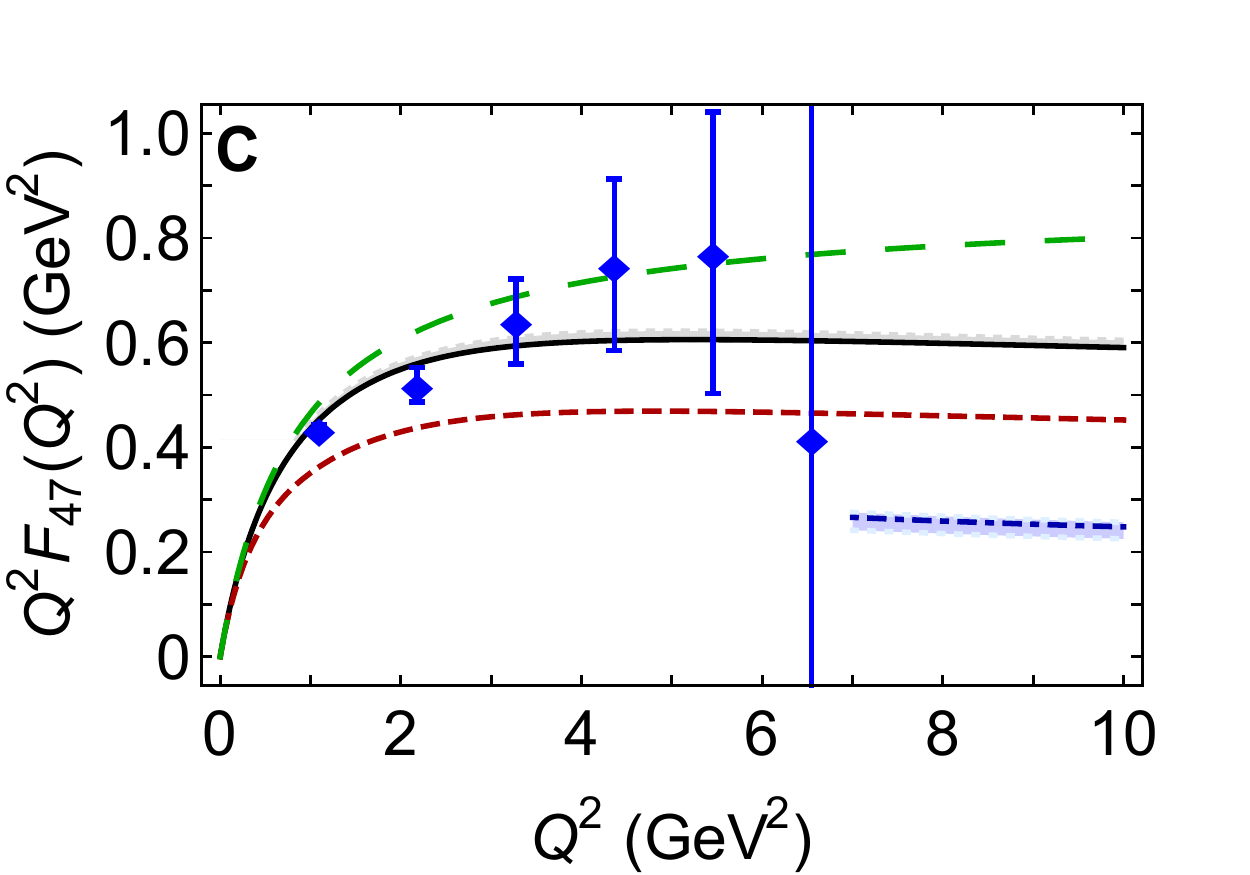}
\includegraphics[width=0.45\textwidth]{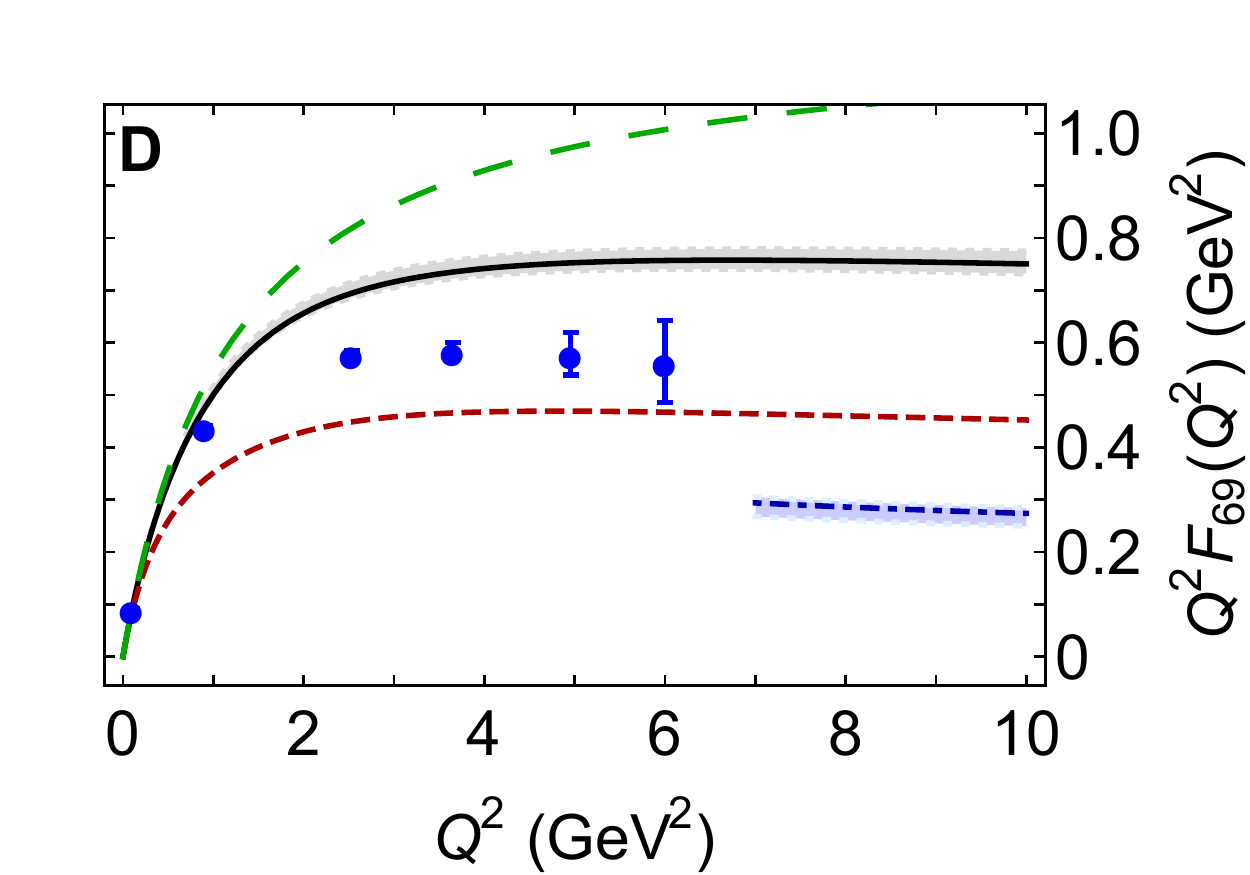}
\caption{\label{FigFFs}
Elastic form factors of pion-like pseudoscalar mesons.
\textbf{A} -- physical pion, $m_\pi = 0.14\,$GeV;
\textbf{B} -- mass-degenerate quarks, meson mass $=0.83\,$GeV;
\textbf{C} -- meson mass  $=0.47\,$GeV; and
\textbf{D} -- meson mass $=0.69\,$GeV.
Curves in each panel.
Solid black curve within grey bands -- our prediction: obtained with $w=0.5\pm0.1\,$GeV in Eq.\,\eqref{defcalG};
long-dashed green curve --  single-pole vector meson dominance result obtained with vector meson mass, $m_V$, computed consistent with the form factor prediction [see Eq.\,\eqref{FitParams}];
and dot-dashed blue curve within blue bands -- result from hard-scattering formula, Eq.\,\eqref{EqHardScattering}, computed with the consistent meson decay constant and PDA.
\textbf{A}.  Dotted purple curve -- Eq.\,\eqref{EqHardScattering} computed with the consistent pion decay constant and conformal-limit PDA, $\varphi^{\rm cl}(x)=6 x (1-x)$; filled-circles and -squares --  data described in Ref.\,\protect\cite{Huber:2008id}; and filled gold diamonds and green triangle -- projected reach and accuracy of forthcoming experiments \protect\cite{E1206101, E12-07-105}.
For comparison, the dashed red curve in the other panels is the black curve from \textbf{A}, \emph{viz}.\ the physical-pion form factor prediction.
\textbf{C} -- filled blue diamonds, lQCD results in Ref.\,\cite{Chambers:2017tuf}; and \textbf{D} -- filled blue circles, lQCD results in Ref.\,\cite{Koponen:2017fvm}.
}
\end{figure*}

\noindent\textbf{3.$\;$Results}.
We have computed the form factors of pion-like mesons at four current-quark masses, corresponding to the physical pion, the lQCD meson masses in Refs.\,\cite{Chambers:2017tuf, Koponen:2017fvm}, and one larger value, obtained by choosing the next evenly-spaced increment in current-quark mass.  The results are reported in Table~\ref{TabResults} and Fig.\,\ref{FigFFs}.

We approached the task without sophistication, using numerical solutions of the relevant gap and Bethe-Salpeter equations to directly evaluate the integral in Eq.\,\eqref{RLFpi}.  Owing to the analytic structure of some of the functions involved \cite{Maris:1997tm, Windisch:2016iud}, this algorithm fails on $Q^2\gtrsim Q_f^2$, where $Q_f^2/{\rm GeV}^2 = 4, 5, 6, 7$, respectively, for each row in Table~\ref{TabResults}.  Ref.\,\cite{Chang:2013nia} solved this problem by using perturbation theory integral representations [PTIRs] \cite{Nakanishi:1969ph} for each matrix-valued function in Eq.\,\eqref{RLFpi}, enabling a reliable computation of the electromagnetic form factor to arbitrarily large-$Q^2$.  Constructing accurate PTIRs is, however, time consuming; and especially so here because one would need to build new PTIRs for each function at every one of the four current-quark masses.  In completing the panels in Fig.\,\ref{FigFFs} we therefore adapted the procedure introduced in Ref.\,\cite{Qin:2017lcd}, assuming that on the displayed domain each form factor can be expressed as
\begin{subequations}
\begin{align}
F_{0^-}(Q^2) & = \frac{1}{1+Q^2/m_V^2} {\mathpzc A}_{0^-}(Q^2)\,,\\
{\mathpzc A}_{0^-}(Q^2) & = \frac{1 + a_1 Q^2 + a_2^2 Q^4}{1+Q^4 (a_2^2/b_u^2) \ln[1+Q^2/\Lambda_{\rm QCD}^2]}\,,
\end{align}
\end{subequations}
where $m_V$ is the appropriate, computed vector meson mass and $a_1, a_2, b_u$ are determined via a least-squares fit to the computed results on $Q^2 \leq Q_f^2$.  The $\omega = 0.5\,$GeV values are (masses in GeV, coefficients in GeV$^{-2}$)
\begin{equation}
\begin{array}{cc|ccc}
m_{0^-} & m_V & a_1 & a_2  & b_u \\\hline
0.14 & 0.77 & -0.14 & 0.50 & 2.12 \\
0.47 & 0.93 & -0.16 & 0.54 & 2.00 \\
0.69 & 1.10 & -0.22 & 0.68 & 1.94 \\
0.83 & 1.21 & -0.22 & 0.81 & 1.89
\end{array}.
\label{FitParams}
\end{equation}
(Empirical values for $m_V/{\rm GeV}$ in rows 1 and 3 are \cite{Olive:2016xmw}: $0.775$, $1.02$.)  We have confirmed this approach is sound by using it to reanalyse the results in Ref.\,\cite{Chang:2013nia}.

The results in Fig.\,\ref{FigFFs}A confirm the analysis in Ref.\,\cite{Chang:2013nia}.  Namely, the calculated  $F_\pi(Q^2)$ agrees semiquantitatively with the prediction of the hard-scattering formula, Eq.\,\eqref{EqHardScattering}, when the PDA appropriate to the empirical scale is used.  The difference between these two curves is explained by a combination of higher-order, higher-twist corrections to Eq.\,\eqref{EqHardScattering} on the one hand and, on the other,  shortcomings in the rainbow-ladder truncation, described above.  Hence, one should expect dominance of hard contributions to the pion form factor for $Q^2\gtrsim 8\,$GeV$^2$.  Notwithstanding this, the normalisation of the form factor is fixed by a pion wave-function whose dilation with respect to $\varphi^{\rm cl}(x)$ is a definitive signature of DCSB. 

In addition to the preceding observations, the panels in Fig.\,\ref{FigFFs} expose numerous features relating to the evolution of these elastic form factors with meson mass.

\begin{enumerate}[label=(\roman*)]
\item The charge radius decreases with increasing mass, \emph{i.e}.\ the bound-states become more pointlike; and $r_{0^-} \propto 1/f_{0^-}$, up to $\ln m_{0^-}$-corrections.  This is illustrated in Fig.\,\ref{rtimef}A and explained elsewhere \cite{Bhagwat:2006xi}.

    $r_{0^-}$ is an intrinsic length-scale in these systems.  The meson becomes a more highly correlated state as it diminishes.  Hence, steadily increasing values of $Q^2$ are required to reach the domain upon which Eq.\,\eqref{EqHardScattering} provides a useful guide to $F_{0^-}(Q^2)$.

\item This last feature is readily apparent in Fig.\,\ref{FigFFs}.  Proceeding anticlockwise from \textbf{A} $\to$ \textbf{C} $\to$ \textbf{D} $\to$ \textbf{B}, the mismatch increases between the direct calculation [solid black curve] and the result obtained using Eq.\,\eqref{EqHardScattering} with the appropriate $f_{0^-}$, $\varphi_{0^-}(x;Q^2)$ [dot-dashed blue curve].

\begin{figure}[t!]
\begin{center}
\includegraphics[clip,width=0.40\textwidth]{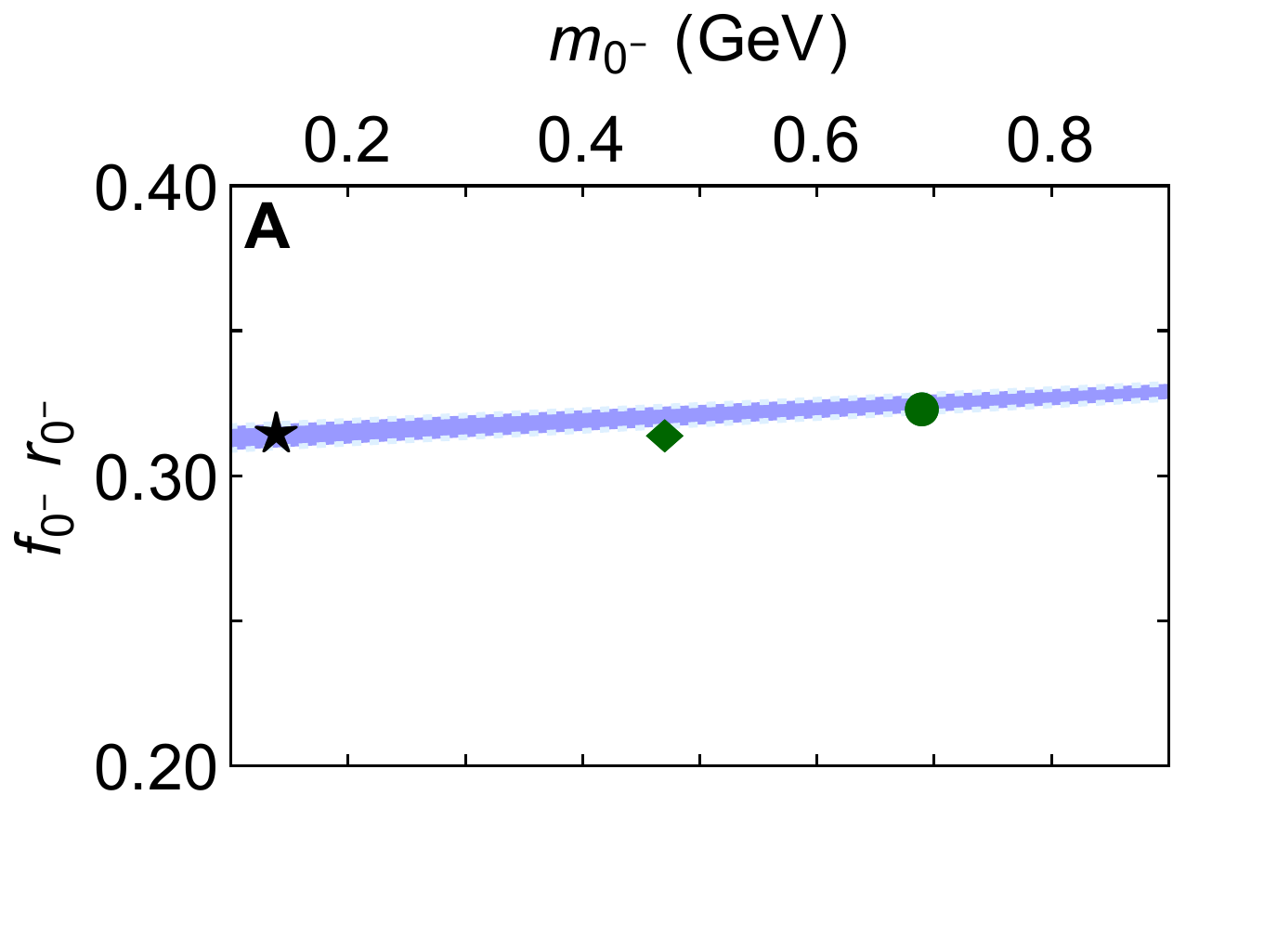}
\vspace*{-9ex}

\includegraphics[clip,width=0.405\textwidth]{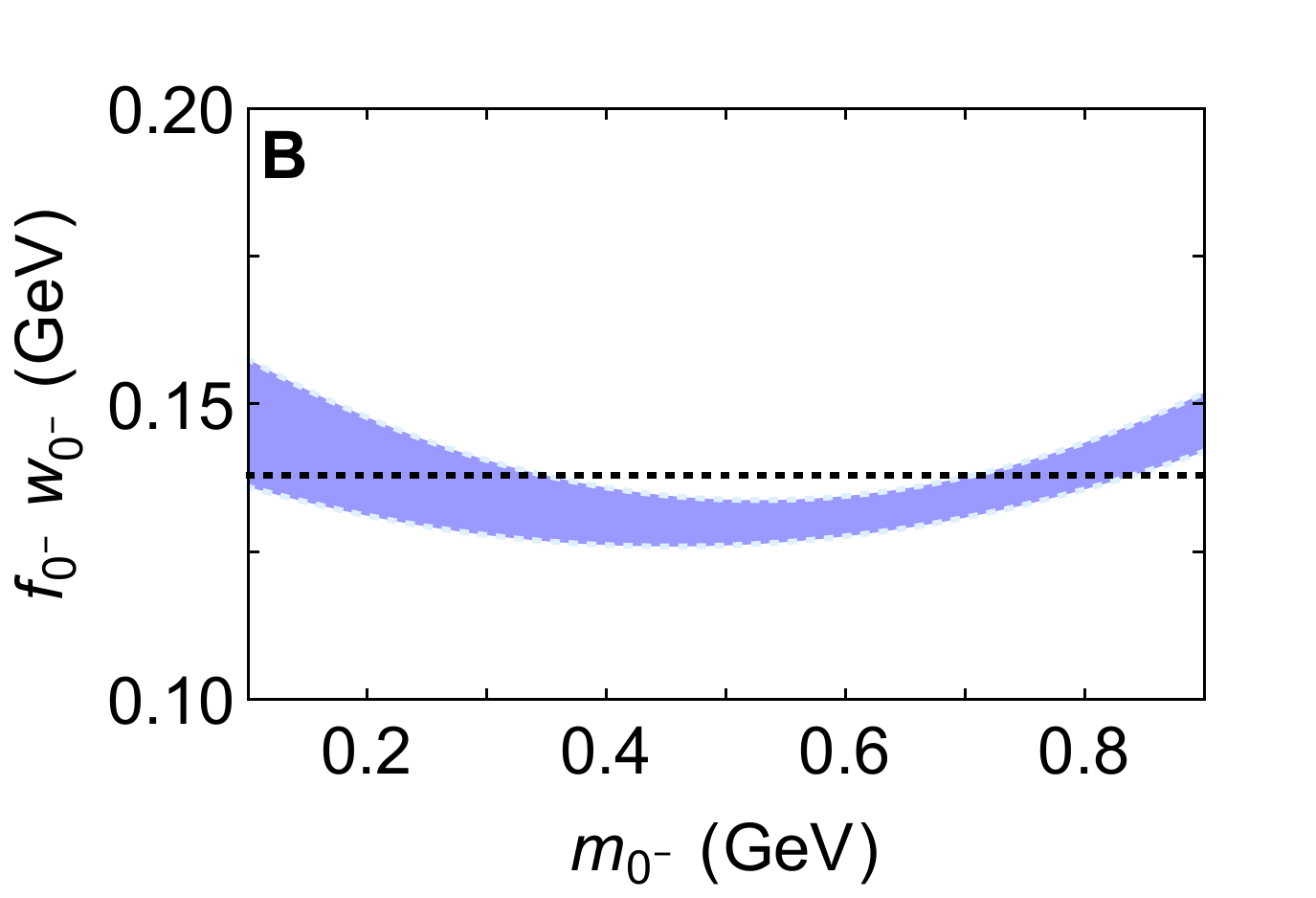}\hspace*{0.5em}
\caption{\label{rtimef}
\textbf{A}. $f_{0^-} r_{0^-}$ as a function of meson mass.  It is nearly constant over a large range \cite{Bhagwat:2006xi, Cloet:2008fw}.  Black star -- empirical value for the pion; green diamond -- lQCD \cite{Chambers:2017tuf}; and green circle -- lQCD \cite{Koponen:2017fvm}.
\textbf{B}. $f_{0^-} w_{0^-}$: $w_{0^-}$ is the $\langle 1/x \rangle$-moment in Eq.\,\eqref{wphi}.  This function takes a minimum value in the neighbourhood of the $s$-quark current-mass and then evolves toward linear growth with $m_{0^-}$, up to logarithmic corrections.  The dotted black line marks the mean value.
The bands in both panels describe the range of results obtained for $\omega\in[0.4,0.6]\,$GeV.
}
\end{center}
\end{figure}

%

\item The failure of the Eq.\,\eqref{EqHardScattering} prediction to increase in magnitude as quickly as the direct calculation is explained by a feature of the meson PDA's $\langle 1/x \rangle$-moment, illustrated in Fig.\,\ref{rtimef}B.  Namely, $f_{0^-} w_{0^-}$ is roughly constant on the domain of meson masses considered: with $\omega=0.5\,$GeV, the integrated relative difference between the computed $m_{0^-}$-dependence and the mean value is just 3\%.  Consequently, the prediction of the hard-scattering formula is weakly varying on $m_{0^-}\in [0.1,0.9]\,$GeV, whereas the form factor itself rises steadily with $m_{0^-}$, owing primarily to the decreasing radius [increasing $f_{0^-}$] of the system.

    Evidently, therefore, the growing Higgs-generated current-quark mass drives away the domain whereupon the exclusive hard-scattering formula is applicable for the associated bound-state.  This property, \emph{viz}.\ that with growing mass, increasingly larger values of $Q_0^2$ are required in order to enter the domain of validity for hard-scattering formulae, is also found in the treatment of $\gamma \gamma^\ast \to\,$neutral-$0^-$-meson transition form factors \cite{Raya:2016yuj}.

    It is worth noting, too, that the minimum of the $f_{0^-} w_{0^-}$ curve occurs in the neighbourhood of the $s$-quark current-mass.  This is a consequence of the fact, shown elsewhere \cite{Ding:2015rkn} and evident from the values of $\alpha_{0^-}$ in Table~\ref{TabResults}, that $\varphi_{0^-}(x;Q^2) \approx \varphi^{\rm cl}(x)$ in this neighbourhood.  With masses increasing away from this domain, $f_{0^-} w_{0^-}$ becomes a linear function, up to $\ln m_{0^-}$-corrections.

\item Notwithstanding these facts, the direct calculation's deviation from the trajectory defined by the single-pole vector-meson-dominance [VMD] prediction [long-dashed green curve] also increases with $m_{0^-}$, and in each case the departure begins at a steadily decreasing value of $Q^2$.  These effects owe to a shift to deeper timelike values of the ground-state vector-meson mass, so that this resonance contribution to the dressed-quark-photon vertex diminishes in importance for the meson-photon coupling, and parallel alterations in the pseudoscalar meson's internal structure.  Such deviation from the VMD prediction is a crucial prerequisite to entering the validity domain of Eq.\,\eqref{EqHardScattering}.

\end{enumerate}

Comparing Figs.\,\ref{FigFFs}B, \ref{FigFFs}C, it seems that the lQCD results in Refs.\,\cite{Chambers:2017tuf, Koponen:2017fvm} are mutually inconsistent: the lighter meson mass in Ref.\,\cite{Chambers:2017tuf} is associated with an elastic form factor which is larger in magnitude than that describing the internal structure of the heavier $0^{-+}$-meson in Ref.\,\cite{Koponen:2017fvm}.  We have insufficient information to resolve this issue; but can observe that whilst the low-scale results from both studies match our predictions, only Ref.\,\cite{Chambers:2017tuf} is consistent with our calculations on the domain of larger-$Q^2$.

\smallskip

\noindent\textbf{4.$\;$Summary and Conclusions}.
%
We employed the leading-order approximation in a symmetry-preserving, continuum analysis of the quark-antiquark bound-state problem to determine electromagnetic form factors of pion-like mesons with masses $m_{0^-}/{\rm GeV}=0.14, 0.47, 0.69, 0.83$ on a spacelike domain that extends to $Q^2 \lesssim 10\,$GeV$^2$; and simultaneously computed the parton distribution amplitudes of each system.
The results exposed an array of novel features, with relevance to experiment and also \emph{ab initio} lattice-QCD studies of these systems.  Of particular significance is the conclusion that the form factor of the physical pion provides the best opportunity for verification of the leading-order, leading-twist factorised hard-scattering formula for such exclusive processes.  This is because the lower bound, $Q_0$, of the domain upon which that formula is valid increases quickly with growing $m_{0^-}$, \emph{i.e}.\ more generally, the inflating mass-scale introduced by increasing Higgs-generated current-quark masses drives away the domain whereupon any relevant exclusive hard-scattering formula is applicable for the associated bound-state.

\smallskip

%
\noindent\textbf{Acknowledgments}.
We are grateful for insightful comments from S.\,J.~Brodsky, R.~Ent, T.~Horn, A.~Lovato and J.~Zanotti.
Work supported by:
the Chinese Government's Thousand Talents Plan for Young Professionals;
the Chinese Ministry of Education, under the \emph{International Distinguished Professor} programme;
and U.S.\ Department of Energy, Office of Science, Office of Nuclear Physics, under contract no.~DE-AC02-06CH11357.


\begin{thebibliography}{10}

\bibitem{Farrar:1979aw}
G.~R. Farrar and D.~R. Jackson,
\newblock Phys. Rev. Lett. {\bf 43}, 246 (1979).

\bibitem{Lepage:1979zb}
G.~P. Lepage and S.~J. Brodsky,
\newblock Phys. Lett. B {\bf 87}, 359 (1979).

\bibitem{Efremov:1979qk}
A.~V. Efremov and A.~V. Radyushkin,
\newblock Phys. Lett. B {\bf 94}, 245 (1980).

\bibitem{Lepage:1980fj}
G.~P. Lepage and S.~J. Brodsky,
\newblock Phys. Rev. D {\bf 22}, 2157 (1980).

\bibitem{Dally:1981ur}
E.~B. Dally {\em et~al.},
\newblock Phys. Rev. D {\bf 24}, 1718 (1981).

\bibitem{Dally:1982zk}
E.~B. Dally {\em et~al.},
\newblock Phys. Rev. Lett. {\bf 48}, 375 (1982).

\bibitem{Amendolia:1984nz}
S.~R. Amendolia {\em et~al.},
\newblock Phys. Lett. B {\bf 146}, 116 (1984).

\bibitem{Amendolia:1986wj}
S.~R. Amendolia {\em et~al.},
\newblock Nucl. Phys. B {\bf 277}, 168 (1986).

\bibitem{Volmer:2000ek}
J.~Volmer {\em et~al.},
\newblock Phys. Rev. Lett. {\bf 86}, 1713 (2001).

\bibitem{Horn:2006tm}
T.~Horn {\em et~al.},
\newblock Phys. Rev. Lett. {\bf 97}, 192001 (2006).

\bibitem{Horn:2007ug}
T.~Horn {\em et~al.},
\newblock Phys. Rev. C {\bf 78}, 058201 (2008).

\bibitem{Huber:2008id}
G.~Huber {\em et~al.},
\newblock Phys. Rev. C {\bf 78}, 045203 (2008).

\bibitem{Blok:2008jy}
H.~P. Blok {\em et~al.},
\newblock Phys. Rev. C {\bf 78}, 045202 (2008).

\bibitem{E12-06-101}
\mbox{Huber, G. M., Gaskell, D.} {\em et~al.},
\newblock \emph{Measurement of the Charged Pion Form Factor to High $Q^2$},
\newblock {a}pproved {J}efferson Lab 12 GeV Experiment E12-06-101, 2006.

\bibitem{E12-07-105}
\mbox{Horn, T., Huber, G. M.} {\em et~al.},
\newblock \emph{Scaling Study of the L/T-Separated Pion Electroproduction Cross
  Section at 11 GeV},
\newblock {a}pproved Jefferson Lab 12 GeV Experiment E12-07-105, 2007.

\bibitem{Horn:2017csb}
T.~Horn,
\newblock EPJ Web Conf. {\bf 137}, 05005 (2017).

\bibitem{Mikhailov:1986be}
S.~Mikhailov and A.~Radyushkin,
\newblock JETP Lett. {\bf 43}, 712 (1986).

\bibitem{Petrov:1998kg}
V.~Y. Petrov, M.~V. Polyakov, R.~Ruskov, C.~Weiss and K.~Goeke,
\newblock Phys. Rev. D {\bf 59}, 114018 (1999).

\bibitem{Brodsky:2006uqa}
S.~J. Brodsky and G.~F. de~Teramond,
\newblock Phys. Rev. Lett. {\bf 96}, 201601 (2006).

\bibitem{Arthur:2010xf}
R.~Arthur {\em et~al.},
\newblock Phys. Rev. D {\bf 83}, 074505 (2011).

\bibitem{Chang:2013pq}
L.~Chang {\em et~al.},
\newblock Phys. Rev. Lett. {\bf 110}, 132001 (2013).

\bibitem{Cloet:2013tta}
I.~C. Clo{\"e}t, L.~Chang, C.~D. Roberts, S.~M. Schmidt and P.~C. Tandy,
\newblock Phys. Rev. Lett. {\bf 111}, 092001 (2013).

\bibitem{Segovia:2013eca}
J.~Segovia {\em et~al.},
\newblock Phys. Lett. B {\bf 731}, 13 (2014).

\bibitem{Braun:2015axa}
V.~M. Braun {\em et~al.},
\newblock Phys. Rev. D {\bf 92}, 014504 (2015).

\bibitem{Horn:2016rip}
T.~Horn and C.~D. Roberts,
\newblock J. Phys. G. {\bf 43}, 073001 (2016).

\bibitem{Zhang:2017bzy}
J.-H. Zhang, J.-W. Chen, X.~Ji, L.~Jin and H.-W. Lin,
\newblock Phys. Rev. D {\bf 95}, 094514 (2017).

\bibitem{Chang:2013nia}
L.~Chang, I.~C. Clo{\"e}t, C.~D. Roberts, S.~M. Schmidt and P.~C. Tandy,
\newblock Phys. Rev. Lett. {\bf 111}, 141802 (2013).

\bibitem{E1207105}
T.~Horn and G.~M. Huber,
\newblock (2007),
\newblock {J}efferson Lab Experiment E12-07-105.

\bibitem{Koponen:2015tkr}
J.~Koponen, F.~Bursa, C.~T.~H. Davies, R.~J. Dowdall and G.~P. Lepage,
\newblock Phys. Rev. D {\bf 93}, 054503 (2016).

\bibitem{Alexandrou:2017blh}
C.~Alexandrou {\em et~al.},
\newblock Phys. Rev. D {\bf 97}, 014508 (2018).

\bibitem{Chambers:2017tuf}
A.~J. Chambers {\em et~al.},
\newblock Phys. Rev. D {\bf 96}, 114509 (2017).

\bibitem{Koponen:2017fvm}
J.~Koponen, A.~C. Zimermmane-Santos, C.~T.~H. Davies, G.~P. Lepage and A.~T.
  Lytle,
\newblock Phys. Rev. D {\bf 96}, 054501 (2017).

\bibitem{Munczek:1994zz}
H.~J. Munczek,
\newblock Phys. Rev. D {\bf 52}, 4736 (1995).

\bibitem{Bender:1996bb}
A.~Bender, C.~D. Roberts and L.~von Smekal,
\newblock Phys. Lett. B {\bf 380}, 7 (1996).

\bibitem{Roberts:1996jxS}
C.~D. Roberts,
\newblock (nucl-th/9609039),
\newblock \emph{Confinement, diquarks and Goldstone's theorem}.

\bibitem{Roberts:1994hh}
C.~D. Roberts,
\newblock Nucl. Phys. A {\bf 605}, 475 (1996).

\bibitem{Maris:1998hc}
P.~Maris and C.~D. Roberts,
\newblock Phys. Rev. C {\bf 58}, 3659 (1998).

\bibitem{Maris:2000sk}
P.~Maris and P.~C. Tandy,
\newblock Phys. Rev. C {\bf 62}, 055204 (2000).

\bibitem{Holl:2005vu}
A.~H{\"o}ll, A.~Krassnigg, P.~Maris, C.~D. Roberts and S.~V. Wright,
\newblock Phys. Rev. C {\bf 71}, 065204 (2005).

\bibitem{Bhagwat:2006pu}
M.~S. Bhagwat and P.~Maris,
\newblock Phys. Rev. C {\bf 77}, 025203 (2008).

\bibitem{Raya:2015gva}
K.~Raya {\em et~al.},
\newblock Phys. Rev. D {\bf 93}, 074017 (2016).

\bibitem{Raya:2016yuj}
K.~Raya, M.~Ding, A.~Bashir, L.~Chang and C.~D. Roberts,
\newblock Phys. Rev. D {\bf 95}, 074014 (2017).

\bibitem{Qin:2011dd}
S.-X. Qin, L.~Chang, Y.-X. Liu, C.~D. Roberts and D.~J. Wilson,
\newblock Phys. Rev. C {\bf 84}, 042202(R) (2011).

\bibitem{Qin:2011xq}
S.-X. Qin, L.~Chang, Y.-X. Liu, C.~D. Roberts and D.~J. Wilson,
\newblock Phys. Rev. C {\bf 85}, 035202 (2012).

\bibitem{Chang:2008ec}
L.~Chang {\em et~al.},
\newblock Phys. Rev. C {\bf 79}, 035209 (2009).

\bibitem{Binosi:2014aea}
D.~Binosi, L.~Chang, J.~Papavassiliou and C.~D. Roberts,
\newblock Phys. Lett. B {\bf 742}, 183 (2015).

\bibitem{Bowman:2004jm}
P.~O. Bowman {\em et~al.},
\newblock Phys. Rev. D {\bf 70}, 034509 (2004).

\bibitem{Boucaud:2011ug}
P.~Boucaud {\em et~al.},
\newblock Few Body Syst. {\bf 53}, 387 (2012).

\bibitem{Ayala:2012pb}
A.~Ayala, A.~Bashir, D.~Binosi, M.~Cristoforetti and J.~Rodr{\'i}guez-Quintero,
\newblock Phys. Rev. D {\bf 86}, 074512 (2012).

\bibitem{Aguilar:2012rz}
A.~Aguilar, D.~Binosi and J.~Papavassiliou,
\newblock Phys. Rev. D {\bf 86}, 014032 (2012).

\bibitem{Binosi:2016xxu}
D.~Binosi, C.~D. Roberts and J.~Rodr{\'i}guez-Quintero,
\newblock Phys. Rev. D {\bf 95}, 114009 (2017).

\bibitem{Binosi:2016nme}
D.~Binosi, C.~Mezrag, J.~Papavassiliou, C.~D. Roberts and
  J.~Rodr{\'i}guez-Quintero,
\newblock Phys. Rev. D {\bf 96}, 054026 (2017).

\bibitem{Gao:2017uox}
F.~Gao, S.-X. Qin, C.~D. Roberts and J.~Rodriguez-Quintero,
\newblock Phys. Rev. D {\bf 97}, 034010 (2018).

\bibitem{Rodriguez-Quintero:2018wma}
J.~Rodr{\'{\i}}guez-Quintero, D.~Binosi, C.~Mezrag, J.~Papavassiliou and C.~D.
  Roberts,
\newblock Few Body Syst. {\bf 59}, 121 (2018).

\bibitem{Skullerud:2003qu}
J.~I. Skullerud, P.~O. Bowman, A.~K{\i}z{\i}lers{\"u}, D.~B. Leinweber and
  A.~G. Williams,
\newblock JHEP {\bf 04}, 047 (2003).

\bibitem{Bhagwat:2004kj}
M.~S. Bhagwat and P.~C. Tandy,
\newblock Phys. Rev. D {\bf 70}, 094039 (2004).

\bibitem{Aguilar:2014lha}
A.~C. Aguilar, D.~Binosi, D.~Iba{\~n}ez and J.~Papavassiliou,
\newblock Phys. Rev. D {\bf 90}, 065027 (2014).

\bibitem{Williams:2015cvx}
R.~Williams, C.~S. Fischer and W.~Heupel,
\newblock Phys. Rev. D {\bf 93}, 034026 (2016).

\bibitem{Binosi:2016rxz}
D.~Binosi, L.~Chang, S.-X. Qin, J.~Papavassiliou and C.~D. Roberts,
\newblock Phys. Rev. D {\bf 93}, 096010 (2016).

\bibitem{Binosi:2016wcx}
D.~Binosi, L.~Chang, J.~Papavassiliou, S.-X. Qin and C.~D. Roberts,
\newblock Phys. Rev. D {\bf 95}, 031501(R) (2017).

\bibitem{Aguilar:2016lbe}
A.~C. Aguilar, J.~C. Cardona, M.~N. Ferreira and J.~Papavassiliou,
\newblock Phys. Rev. D {\bf 96}, 014029 (2017).

\bibitem{Bermudez:2017bpx}
R.~Bermudez, L.~Albino, L.~X. Guti{\'e}rrez-Guerrero, M.~E. Tejeda-Yeomans and
  A.~Bashir,
\newblock Phys. Rev. D {\bf 95}, 034041 (2017).

\bibitem{Cyrol:2017ewj}
A.~K. Cyrol, M.~Mitter, J.~M. Pawlowski and N.~Strodthoff,
\newblock Phys. Rev. D {\bf 97}, 054006 (2018).

\bibitem{Eichmann:2008ef}
G.~Eichmann, I.~C. Clo{\"e}t, R.~Alkofer, A.~Krassnigg and C.~D. Roberts,
\newblock Phys. Rev. C {\bf 79}, 012202(R) (2009).

\bibitem{Eichmann:2012zz}
G.~Eichmann,
\newblock Prog. Part. Nucl. Phys. {\bf 67}, 234 (2012).

\bibitem{Olive:2016xmw}
C.~Patrignani {\em et~al.},
\newblock Chin. Phys. C {\bf 40}, 100001 (2016).

\bibitem{Bornyakov:2016dzn}
V.~G. Bornyakov {\em et~al.},
\newblock Phys. Lett. B {\bf 767}, 366 (2017).

\bibitem{Maris:1997tm}
P.~Maris and C.~D. Roberts,
\newblock Phys. Rev. C {\bf 56}, 3369 (1997).

\bibitem{Krassnigg:2009gd}
A.~Krassnigg,
\newblock PoS {\bf CONFINEMENT8}, 075 (2008).

\bibitem{Maris:1999bh}
P.~Maris and P.~C. Tandy,
\newblock Phys. Rev. C {\bf 61}, 045202 (2000).

\bibitem{Li:2016dzv}
B.~L. Li {\em et~al.},
\newblock Phys. Rev. D {\bf 93}, 114033 (2016).

\bibitem{Windisch:2016iud}
A.~Windisch,
\newblock Phys. Rev. C {\bf 95}, 045204 (2017).

\bibitem{Nakanishi:1969ph}
N.~Nakanishi,
\newblock Prog. Theor. Phys. Suppl. {\bf 43}, 1 (1969).

\bibitem{Qin:2017lcd}
S.-X. Qin, C.~Chen, C.~Mezrag and C.~D. Roberts,
\newblock Phys. Rev. C {\bf 97}, 015203 (2018).

\bibitem{E1206101}
\mbox{Huber, G. M. and Gaskell, D.} {\em et~al.},
\newblock (2006),
\newblock {J}efferson Lab Experiment E12-06-101.

\bibitem{Bhagwat:2006xi}
M.~S. Bhagwat, A.~Krassnigg, P.~Maris and C.~D. Roberts,
\newblock Eur. Phys. J. A {\bf 31}, 630 (2007).

\bibitem{Cloet:2008fw}
I.~C. Clo{\"e}t and C.~D. Roberts,
\newblock PoS {\bf LC2008}, 047 (2008).

\bibitem{Ding:2015rkn}
M.~Ding, F.~Gao, L.~Chang, Y.-X. Liu and C.~D. Roberts,
\newblock Phys. Lett. B {\bf 753}, 330 (2016).

\end{thebibliography}

\end{document}